\documentclass[aps, prb,twocolumn,floatfix,superscriptaddress]{revtex4}

\usepackage{graphicx}
\usepackage{bbm}
\usepackage{amsmath, amssymb}

\usepackage{hyperref}
\hypersetup{colorlinks=true, urlcolor=blue, linkcolor=blue, citecolor=blue}

\begin{document}
\title{Fano Resonances in Mismatched C$_3$N Nanoribbon Junctions}

\author{Andor Che Papior}
\affiliation{Faculty of Physics, Technische Universit\"at Dresden, 01062 Dresden, Germany}
\author{Van-Truong Tran}
\affiliation{Université Paris-Saclay, Centre de Nanosciences et de Nanotechnologies, 91120, Palaiseau, France}
\author{Roberto D'Agosta}
\affiliation{Nano-Bio Spectroscopy Group and European Theoretical Spectroscopy Facility (ETSF), Departamento de Pol\'imeros y Materiales Avanzados: F\'isica, Qu\'imica y Tecnolog\'ia, Universidad del Pa\'is Vasco UPV/EHU, Avenida de Tolosa 72, E-20018 San Sebasti\'an, Spain}
\author{Stefan Kurth}
\email[Corresponding author: ]{stefan.kurth@ehu.es}
\affiliation{Nano-Bio Spectroscopy Group and European Theoretical Spectroscopy Facility (ETSF), Departamento de Pol\'imeros y Materiales Avanzados: F\'isica, Qu\'imica y Tecnolog\'ia, Universidad del Pa\'is Vasco UPV/EHU, Avenida de Tolosa 72, E-20018 San Sebasti\'an, Spain}
\affiliation{IKERBASQUE, Basque Foundation for Science, Plaza Euskadi 5, E-48009 Bilbao, Spain}
\affiliation{Donostia International Physics Center (DIPC), Paseo Manuel de Lardizabal 4, 
  E-20018 San Sebasti\'{a}n, Spain}

\date{\today}

\begin{abstract}
Mismatched junctions formed by two C$_3$N zigzag nanoribbons of different widths provide a useful setting for studying quantum interference effects involving edge state transport. A crucial ingredient for this interference to appear is, besides the presence of edge states, the formation of localized interface states at the mismatched interface of the junction. 
At the level of a tight-binding model it is shown that, by means of an external gate potential, one of the edge state energy bands can selectively be shifted into the energy range of the localized interface states. 
The resulting coupling between the edge and localized interface states gives rise to pronounced Fano resonances 
in both the density of states and the transmission spectrum with line shapes well described by the canonical Fano formula. Furthermore, it is found that the geometrical mismatch of the junction not only determines the number of resonances but also the energetic orientation of their asymmetric line shapes. 
These results identify mismatched C$_3$N nanojunctions as a tunable and robust platform for engineering interference-driven transport.  
\end{abstract}

\maketitle

\section{Introduction}

Ever since the experimental exfoliation of graphene \cite{NovoselovGeim:04}, atomically thin two-dimensional (2D) materials have attracted enormous interest stemming not only from their unique physical properties in extended layers but also from the possibility of assembling diverse nanostructures with well-controlled functionalities. For instance, a research direction that has gained significant momentum explores vertical stacking of different 2D materials with similar lattice constants in order to reduce local strain and structural deformations \cite{Geim2013vdW}, aiming at devices exhibiting phenomena not present in the isolated constituents, such as superconductivity in twisted bilayer graphene \cite{Geim2013vdW,LopesdosSantos2007,Bistritzer2011,Cao2018CI,Cao2018SC}. 

Beyond vertical stacking, an alternative and highly versatile route to engineering the properties of 2D materials consists of laterally confining them into nanoribbons.  
In particular, nanoribbons built from hexagonal unit cells and terminated with either zigzag or armchair edges are among the most common graphene nanostructures that can be synthesized directly \cite{Ruffieux2016} allowing for a high degree of control over features such as edge termination and chemical substitution. Reflection-symmetric zigzag graphene nanoribbons typically exhibit localized edge states that are antiferromagnetically coupled \cite{Son2006}; however, chemical modification of one edge can lead to states localized on only one side of the ribbon \cite{Song_etal:25}.

Going beyond graphene, edge states can be observed in other nanoribbons as well.
Indeed, it has been shown  that nanoribbons corresponding to honeycomb monolayers with chemical composition A$_3$B can have edge states on one or both edges of the ribbon, depending on the precise configurations \cite{KamedaLiuDuttaWakabayashi:19}. Moreover, these edge states--ideally suited to control electronic transport--are topological, i.e., they are robust against external perturbations. 

A particular, experimentally realizable example of the A$_3$B structure mentioned above, is polyaniline (C$_3$N) \cite{Mahmood_etal:16,Yang_etal:17}. In bulk, polyaniline is a conductive polymer commonly used in sensors, supercapacitors, rechargeable batteries, corrosion-resistant coatings, and, more recently, as thermoelectric material \cite{SNOOK20111,Caballero2017}. In the 2D monolayer, C$_3$N forms a honeycomb structure in which one in every four carbon atoms is replaced by nitrogen. The resulting planar lattice preserves the $sp^2$ bonding network characteristic of graphene while breaking the equivalence of the two sublattices. Monolayer C$_3$N is a semiconductor with an indirect band gap of 
$\simeq$0.38-1.09 eV from Density Functional Theory (DFT) calculations, depending on the chosen method to parametrize the exchange-correlation energy \cite{Chen2019}.  Experimental values are reported at 2.37 eV \cite{Mahmood_etal:16} for monolayer and about 0.9 eV for bilayer C$_3$N due to interlayer
hybridization \cite{Wei2021}. Furthermore, monolayer C$_3$N exhibits high
mechanical stability together with relatively large thermal conductivity
\cite{Jiao2021,Zhou2017,Chen2020}. Owing to this combination of structural
similarity to graphene and distinct electronic properties, C$_3$N has
attracted considerable attention as a promising platform for nanoelectronic
and optoelectronic applications.

For nanoribbons in 1D, previous DFT calculations \cite{metalandsemiconductor} show that, according to the edge termination and the width of the ribbon, one can find a transition from direct band semiconductor to a metal where the Fermi energy crosses one or two states 
localized at the nanoribbon edges. The effect of edge passivation on electronic transport has been studied in Ref.~\onlinecite{Xia_etal:19}. Furthermore, mismatched nanojunctions formed by C$_3$N nanoribbons of different widths have been studied in Ref.~\onlinecite{HeGuoYanZeng:18} where the focus has been on negative differential conductance. 

In the present work, we highlight another possible effect of nanostructuring: for mismatched nanojunctions composed of two C$_3$N zigzag nanoribbons (ZNR) with different widths, we can control the formation of localized interface or surface states which, by suitable electrostatic potentials, can be  brought to interact/hybridize with the edge states of the ideal periodic ribbon. Since the edge states of the nanoribbon belong to an energy band with a continuum of energies while the surface states are localized (and thus correspond to discrete energies of the system), the interplay between localized and continuum states leads to the formation of Fano resonances \cite{Fano:61} which are well reflected in both
the density of states (DOS) and the transmission function. Fano resonances
have been discussed for applications in nanoplasmonics 
\cite{Lukyanchuk_etal_2010} and metamaterials
\cite{Lukyanchuk_etal_2010,He_etal_nanoph_2025}.

The remainder of the paper is organized as follows: in Sec.~\ref{sec:model} we briefly introduce both the model and the techniques used to investigate mismatched C$_3$N nanojunctions. In order to point out the different ingredients necessary for the formation of Fano resonances in these structures, we proceed step by step: we start with ideal, periodic ZNRs (Sec.~\ref{sec:periodic}) to highlight the presence and control of edge states. In addition to these edge states of the ideal periodic ribbon, {\em semi-infinite} ZNRs (Sec.~\ref{sec:semi_infinite}) also feature localized interface (or surface) states at the newly created surface. Putting everything together, suitably designed mismatched nanojunctions (Sec.~\ref{sec:mismatched_junctions}) allow for the appearance of Fano resonances in the transmission function and the density of states due to the interaction between the localized surface states and the band of edge states. Finally, in Sec.~\ref{sec:conclusions} we summarize and present our conclusions.

\section{Model and methodology}
\label{sec:model}

\begin{figure}[tb]
    \includegraphics[width=1.0\linewidth]{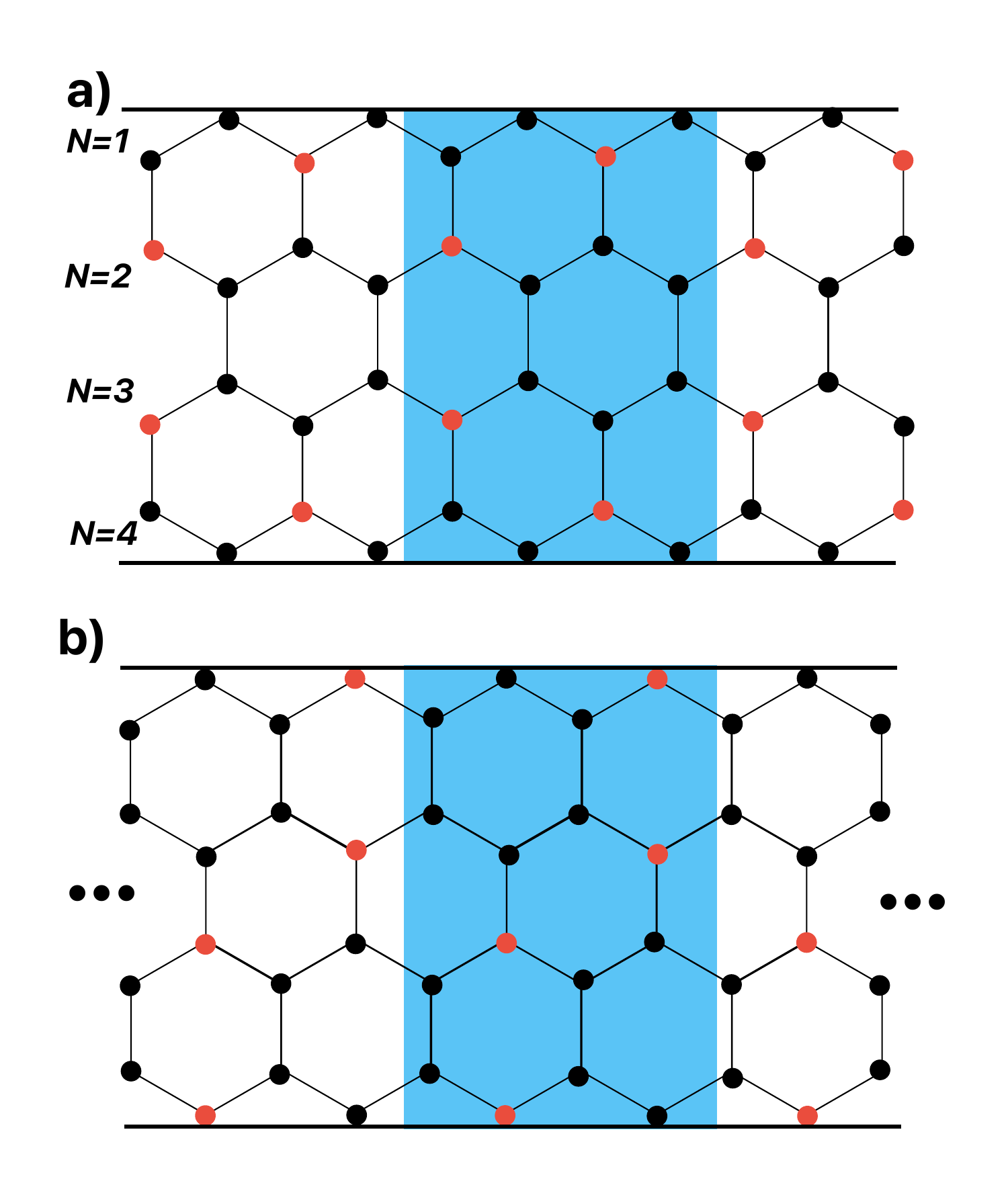}
    \caption{Unit cell (highlighted) of the ideal C$_3$N zigzag nanoribbon (ZNR) for $N_T=4$ atomic chains and two different edge terminations. For any even $N_T$, these are the only possible edge terminations of the ideal ZNR. For the termination a) (upper panel), all the edge atoms are carbons while for termination b) (lower panel) both carbon and nitrogen atoms form the edge.} 
    \label{fig:pat1pat2}
\end{figure}

We studied different geometric configurations of C$_3$N ZNRs, as shown in Fig. \ref{fig:pat1pat2}. The geometric configurations differ in the atomic sequence at the edges of the ribbon: while for the configuration of Fig.~\ref{fig:pat1pat2} a) the edge atoms, for the top and bottom edges, are all carbon (C-C edge termination),
%. On the other hand, 
in the configuration of Fig. \ref{fig:pat1pat2} b) both carbon and nitrogen atoms form the edges (C-N edge termination). The width of a ZNR is determined by the total number of atomic chains, $N_T$, composing the ribbon. For even $N_T$, both the upper and lower edges of the ZNR have the {\em same} edge termination (C-C or C-N, see Fig.~\ref{fig:pat1pat2}).  Note that if $N_T/2$ is even,
the ribbon is mirror symmetric with respect to its central axis, while if $N_T/2$ is odd, the system is invariant under the mirror operation at the central axis followed by translation of half a unit cell in the direction of the ribbon (glide mirror symmetry). On the other hand, for $N_T$ odd, one of the edges has C-C, the other one C-N termination, and the point symmetry is broken.

To investigate the electronic structure and quantum transport in mismatched
C$_3$N nanojunctions, we employed a tight-binding (TB) model in combination with
the Non-Equilibrium Green’s Functions (NEGF) formalism. The TB Hamiltonian is
given as \cite{CastroNeto_etal_2009,tightbindingparameter}
\begin{equation}
\hat{H} = \sum_{i,\sigma}\left( \varepsilon _i + v_i \right) \hat{c}_{i\sigma}^{\dagger} \hat{c}_{i \sigma}   - \sum_{\{i,j\},\sigma}t_{ij} \left( \hat{c}_{i \sigma}^{\dagger} \hat{c}_{j \sigma} + H.c. \right),
\label{eq_one}
\end{equation}
where $\hat{c}_{i \sigma}^{\dagger}$ ($\hat{c}_{i \sigma}$) are the creation (annihilation) operators for an electron with spin $\sigma$ at site $i$. In the following, all parameters are taken to be independent of spin $\sigma$, i.e., we study non-magnetic ribbons only. The $\varepsilon _i$ are the on-site energies at site $i$ (which depend on the atomic species at this site),  $v_i$ is an additional electrostatic gate potential,  and $t_{ij}$ refers to the hopping  parameter between atoms at the $i$-th and $j$-th sites. The double sum in the second term of Eq.~(\ref{eq_one}) runs only over nearest neighbor sites and the corresponding hopping matrix element $t_{ij}=t$ was taken to be the same for both C-C and C-N bonds. In this work, we used $t=3.1$ eV for the hopping while the on-site energies are $\varepsilon_C=-2.6$ eV for the carbon atoms and $\varepsilon_N=-4.73$ eV for the nitrogen atoms. These values were determined \cite{tightbindingparameter} by comparing the tight-binding bands with those obtained from Density Functional Theory (DFT) using the HSE functional \cite{HeydScuseriaErnzerhof:03}. Although the employed TB model was initially fitted to the electronic structure of 2D C$_3$N sheets, the well-established success of similar approaches in graphene \cite{CastroNeto_etal_2009} suggests that this model still provides a reliable framework for examining the qualitative electronic properties and quantum effects in C$_3$N nanostructures.

The electronic structure of the ideal C$_3$N ZNRs is calculated by diagonalizing the matrix obtained from Eq.~(\ref{eq_one}) for a unit cell with periodic boundary conditions in one dimension. In the case of open systems where a central device region is connected to a left (L) and right (R) lead, we adopted an embedding scheme to evaluate the retarded Green's function ${\bf G}_C$ of the defined central region as \cite{Datta:95,Datta:05,LewenkopfMucciolo:13}
\begin{equation}
    {\bf G}_C(E) = %\left((E +i\eta)\mathbbm{1}-{\bf H}_C-{\bf
    \left((E+i \eta)\mathbbm{1}-{\bf H}_C-{\bf\Sigma}_L-{\bf \Sigma}_R\right)^{-1}
\end{equation}
where $\mathbbm{1}$ is the unit matrix, ${\bf H}_C$ is the Hamiltonian matrix describing the central region, ${\bf \Sigma}_{\alpha}(E)$ ($\alpha=L,R$) are the (retarded) embedding self energies of lead $\alpha$, and $\eta$ is a positive infinitesimal. A lead is modeled as a semi-infinite ZNR and the corresponding embedding self energy is calculated with the Sancho-Rubio iterative technique \cite{LopezSanchoLopezSanchoRubio:85}. The local density of states (LDOS) $D_j(E)$ at site $j$ of the central region is computed as
\begin{equation}
    D_j(E) = -\frac{1}{\pi} {\rm Im}\left( {\bf G}_C(E)\right)_{jj}
\end{equation}
while the transmission function through the central region is given by
\begin{equation}
    T(E) = \operatorname{Tr}[{\bf \Gamma}_L{\bf G}_C {\bf \Gamma}_R {\bf G}_C^{\dagger}]
\end{equation}
where ${\bf \Gamma}_{\alpha} = i ({\bf \Sigma}_{\alpha} - {\bf \Sigma}_{\alpha}^{\dagger})$. 

\section{Periodic ribbons}
\label{sec:periodic}

We first examine and analyze the band structures of periodic ZNRs to lay the foundations for an understanding of the transport properties.

\begin{figure}[tb]
    \includegraphics[width=1.\linewidth]{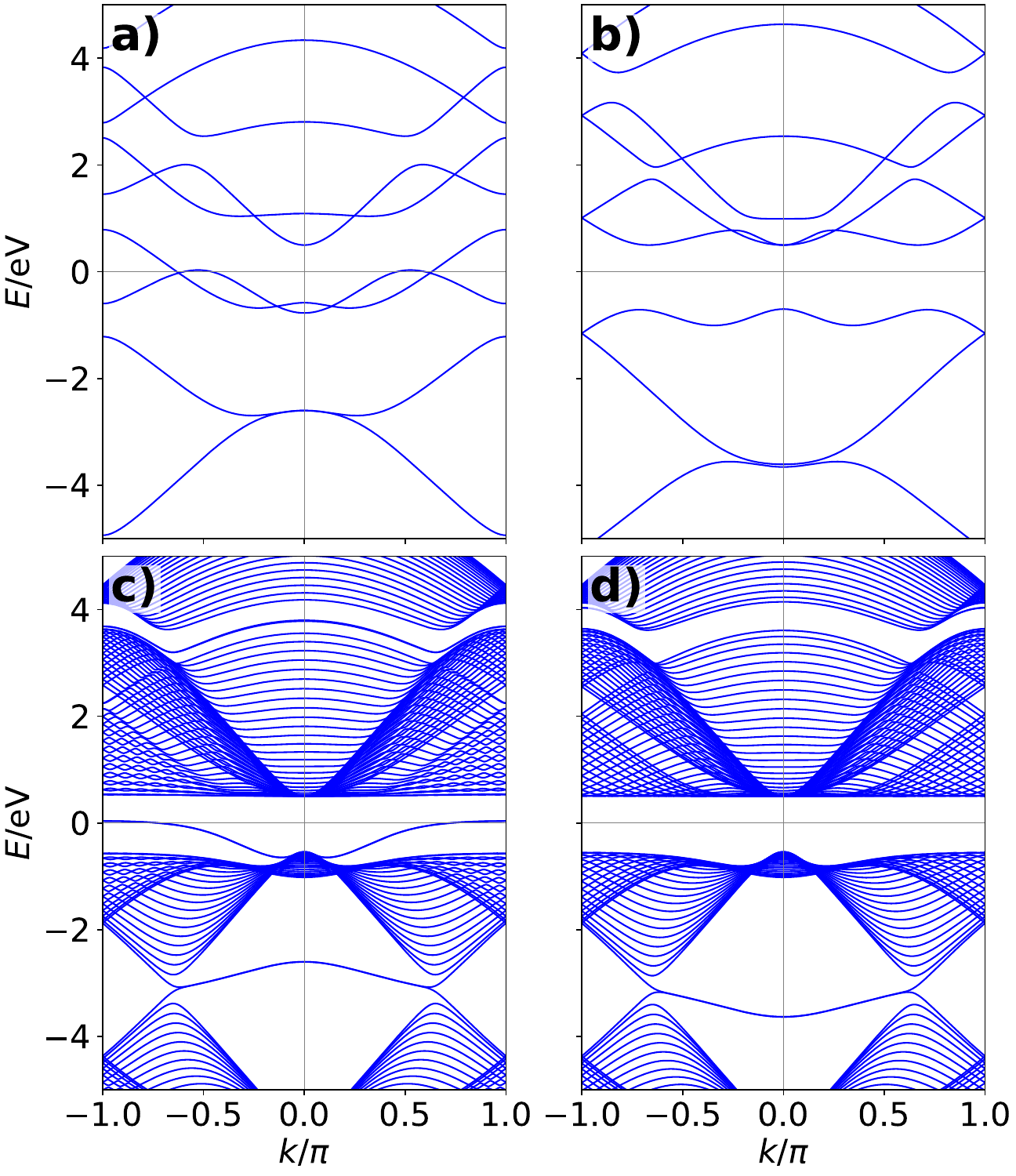}
        \caption{Band structure of C$_3$N ZNRs for different $N_T$. a) $N_T=4$ and C-C edge termination, b) $N_T=4$ and C-N edge, c) $N_T=50$ for C-C edge and d) $N_T=50$ for C-N edge.}
        \label{fig:bands}
\end{figure}

In Fig.~\ref{fig:bands}, we show the band structure for two ideal C$_3$N ZNRs
of different widths, $N_T=4$ and $N_T=50$, for the two edge terminations shown
in Fig.~\ref{fig:pat1pat2}. The edge termination plays a crucial role in the
electronic structure, i.e., for the narrow ZNR with $N_T=4$ we see that the
electronic structure corresponding to the C-N edge termination
(Fig.~\ref{fig:bands} b)) exhibits a well developed band gap around the Fermi
energy $\varepsilon_F=0$ eV which is absent in the case of the C-C termination
(Fig.~\ref{fig:bands} a)). For the wider ZNRs (with even $N_T$), the band gap
for the C-N termination is confirmed in Fig.~\ref{fig:bands} d). These
findings are consistent with the results of DFT calculations for C$_3$N ZNRs
in \cite{metalandsemiconductor}. For the C-C edge termination and $N_T=50$, as
shown in Fig.~\ref{fig:bands} c), there are two almost degenerate bands close
to $\varepsilon_F$ . Closer inspection reveals that these bands correspond to
edge states of the ZNR, i.e., the states localize at the edges of the ribbon.
We find that, close to the Fermi energy and  
for arbitrary width $N_T$ of the ZNR, edge states are only present for edge terminations without
nitrogen atoms in edge positions \cite{Berdiyorov_2024,metalandsemiconductor}.
Consequently, for even $N_T$ with C-C edge termination, the ZNRs always
support two edge states, while those with odd $N_T$ have only a single one.
In the following sections, we focus exclusively on ZNR structures that host a
pair of edge states. 

\begin{figure}[tb]
  \includegraphics[width=1.0\linewidth]{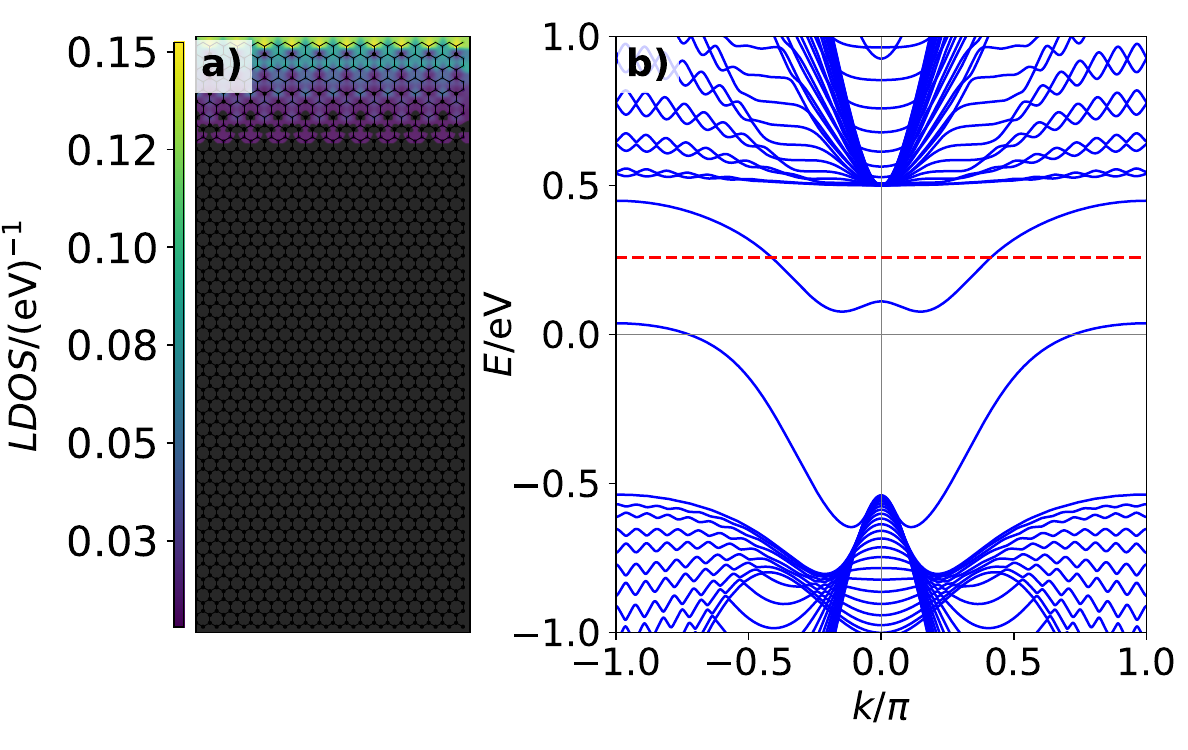}
  \caption{a) LDOS of the shifted band at $E=0.0258$ eV indicated by the
    dashed red line in panel b) and b) band structure of the periodic $N_T = 50$ ZNR with C-C edge termination. A constant  electrostatic potential $v_G=0.8$ eV is added to all sites of the first two chains.}
    \label{fig:bandsbrokensymm}
\end{figure}

Since the edge states are localized in the vicinity of the ribbon edges and the corresponding bands are energetically well separated from the other bands, it is possible to energetically shift a single edge band by applying a constant electrostatic gate potential $v_i$ only to the sites close to the edge. For instance, in Fig.~\ref{fig:bandsbrokensymm} we show the band structure for an ideal ZNR of width $N_T=50$ and C-C edge termination with an additional gate potential  $v_G=0.8$ eV applied to all sites on the chains $N=1$ and $N=2$. This additional gate potential clearly lifts the quasi-degeneracy of the two edge-state bands, resulting in two distinct bands near $\varepsilon_F$. As expected, the higher-energy band is associated with the edge state localized at the upper boundary (see the LDOS panel in Fig.~\ref{fig:bandsbrokensymm}), while the lower-energy band corresponds to the state at the lower edge of the ribbon. We have checked (not shown) that application of the 
gate potential mostly affects the energies of the edge bands while bands with states of bulk character change only little. 
For small gates the energetic shift of the edge bands is approximately linear but for larger values of $v_G$ the shift 
saturates at the lower energy of the bulk state bands.

We note in passing that for ribbons with odd $N_T$, we can apply the same approach to shift the (single) edge state band in the system.

\section{Semi-infinite nanoribbons}
\label{sec:semi_infinite}

As a first step towards understanding and engineering the properties of a mismatched nanojunction, we aim at understanding the electronic structure of a {\em semi-infinite} ZNR, i.e., we cut the ideal ZNR and thus create another edge perpendicular to the edges of the original ZNR. This additional edge has armchair structure. 
Since the unit cell of the ribbon contains four vertical columns of atoms, there are four distinct configurations for defining the interface, depending on the position where the ZNR is cut. Among these, only the two configurations for which  both carbon and nitrogen atoms comprise the armchair interface are capable of hosting interface states within the energy range of 0 to 0.5 eV, exactly the energy window to which edge states can be shifted via application of the gate potential \cite{DingZhangLiZhou:24}.

\begin{figure}[tb]
    \includegraphics[width=1.0\linewidth]{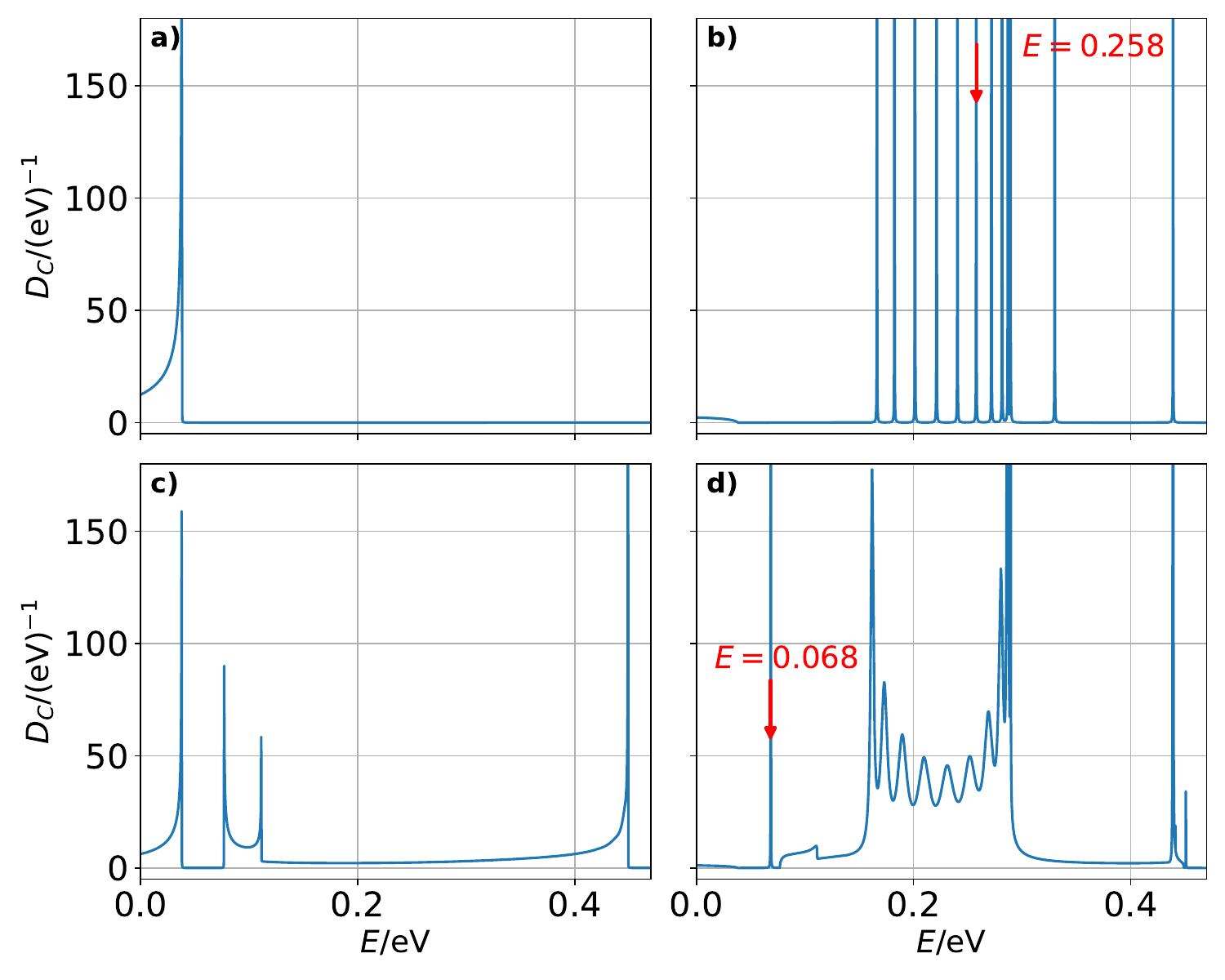}
    \caption{Density of states of central region $D_C(E)$ (for definition, see main text) for the $N_T=50$ ZNR with C-C edge termination for different configurations. a) ideal periodic ZNR with no extra electrostatic potential. b) Semi-infinite ZNR without additional potential. The sharp peaks correspond to localized surface states formed at the armchair termination of the semi-infinite ZNR. c) ideal periodic ZNR with additional gate potential $v_G=0.8$ eV for all atoms of chains $N=1$ and $N=2$. d) Semi-infinite ZNR with gate potential as in c). Many of the localized states found in b) now hybridize with the energy band of the upper edge state of the ZNR. The remaining sharp peaks correspond to localized corner states. The LDOS corresponding to the two highlighted energies in panels b) and d) are shown in Fig.~\ref{fig:interface_corner_state}. }
    \label{fig:DOScomparison}
\end{figure}
\begin{figure}[t!]
    \includegraphics[width=0.8\linewidth]{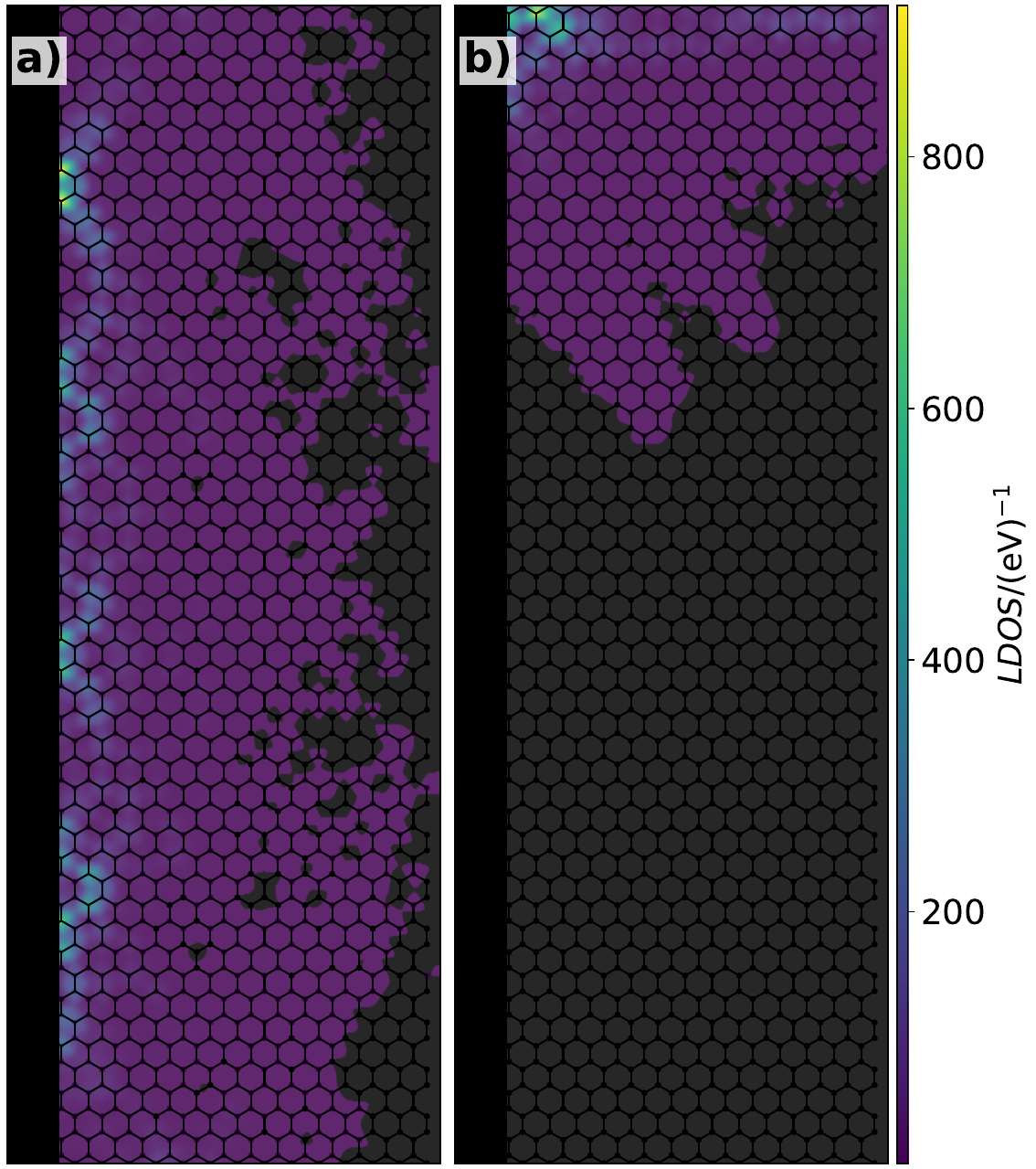}
    \caption{Left panel: LDOS of a semi-infinite ZNR of width $N_T = 50$ without gate potential at $E = 0.258$ eV (the peak marked by an arrow in Fig.~\ref{fig:DOScomparison} b)) corresponding to a surface state. Right panel: LDOS of the semi-infinite $N_T=50$ ZNR with applied gate potential for energy $E=0.068$ eV (the peak marked by an arrow in Fig.~\ref{fig:DOScomparison} d)) corresponding to a corner state. 
    For both panels, the visualized region consists of $m=16$  unit cells  and the hard wall boundary of the semi-infinite ZNRs is indicated by the black region at the left side of the figures.}
    \label{fig:interface_corner_state}
\end{figure}

To highlight the differences in the electronic structure of periodic and semi-infinite ribbons, we computed the corresponding density of states (DOS),
both without and with the application of an edge potential, see Fig.~\ref{fig:DOScomparison}. Here the DOS of the central region is evaluated as $D_C(E) = \sum_j D_j(E)$ with the sum running over all sites of the central region.
We considered ribbons of width $N_T=50$ and C-C edge termination. In panel Fig.~\ref{fig:DOScomparison} a), $D_C$ is shown for the ideal, periodic ZNR in the absence of an applied gate potential while panel b) of the same figure displays the $D_C$ of the corresponding {\em semi-infinite} ZNR again without addition of the gate potential. The results in both panels differ significantly: panel (a) shows a typical 1D van Hove singularity, while panel (b) reveals a series of discrete peaks. This emergence of discrete energy levels points to the formation of localized states caused by the introduction of the additional armchair edge termination of the ZNR. This is confirmed by inspection of the LDOS, as illustrated in Fig.~\ref{fig:interface_corner_state} a). These states live in an energy range where no continuum (band) states are available. As we 
have seen above, adding a localized gate potential to the edge of the ribbons allows shifting the band of the edge state. In Fig.~\ref{fig:DOScomparison} c) we show the $D_C$ of the ideal, periodic ZNR with the gate potential $v_G=0.8$ eV added to all sites of the $N=1$ and $N=2$ chains. A clear shift of the edge-state band can be observed. If we now look at the $D_C$ of Fig.~\ref{fig:DOScomparison} d) for the {\em semi-infinite} ZNR with this gate potential added to all atoms on the same $N=1$ and $N=2$ chains, the energy of the localized states now is in the range of the shifted edge band and can thus hybridize. This leads to the formerly sharp peaks still being visible but now broadened. There are a few remaining sharp peaks below and above the band which correspond to localized corner states. 

In panel a) of Fig.~\ref{fig:interface_corner_state} we show the localized interface state of 
the semi-infinite ZNR without additional gate potential at  energy $E=0.258$ eV (marked by the red arrow in Fig.~\ref{fig:DOScomparison} b)) which is clearly localized at the armchair interface. 
Panel b) of the same figure, on the other hand, shows the state at energy
$E=0.068$ eV marked in Fig.~\ref{fig:DOScomparison} d) which is clearly localized in a corner of the semi-infinite ZNR with gate potential applied to the chains $1$ and $2$. We note that for the surface state of Fig.~\ref{fig:interface_corner_state} a), the LDOS maxima always coincide with the positions of the carbon atoms. On the other hand, the corner state of panel b) also shows LDOS maxima localized on nitrogen atoms.

\section{Mismatched nanojunctions}
\label{sec:mismatched_junctions}

In the present section, we concentrate on quantum transport in nanojunctions
formed by two semi-infinite ZNR leads of different widths. Similar mismatched
C$_3$N nanojunctions have been studied in Ref.~\onlinecite{HeGuoYanZeng:18} with a focus on effects of electronic transport such as negative differential resistance or rectification. 
Here we focus on a different aspect, the formation of Fano resonances \cite{Fano:61}, reflected in both the DOS and the transmission function. 

\subsection{Engineering mismatched nanojunctions}

\begin{figure}[tb]
    \includegraphics[width=1\linewidth]{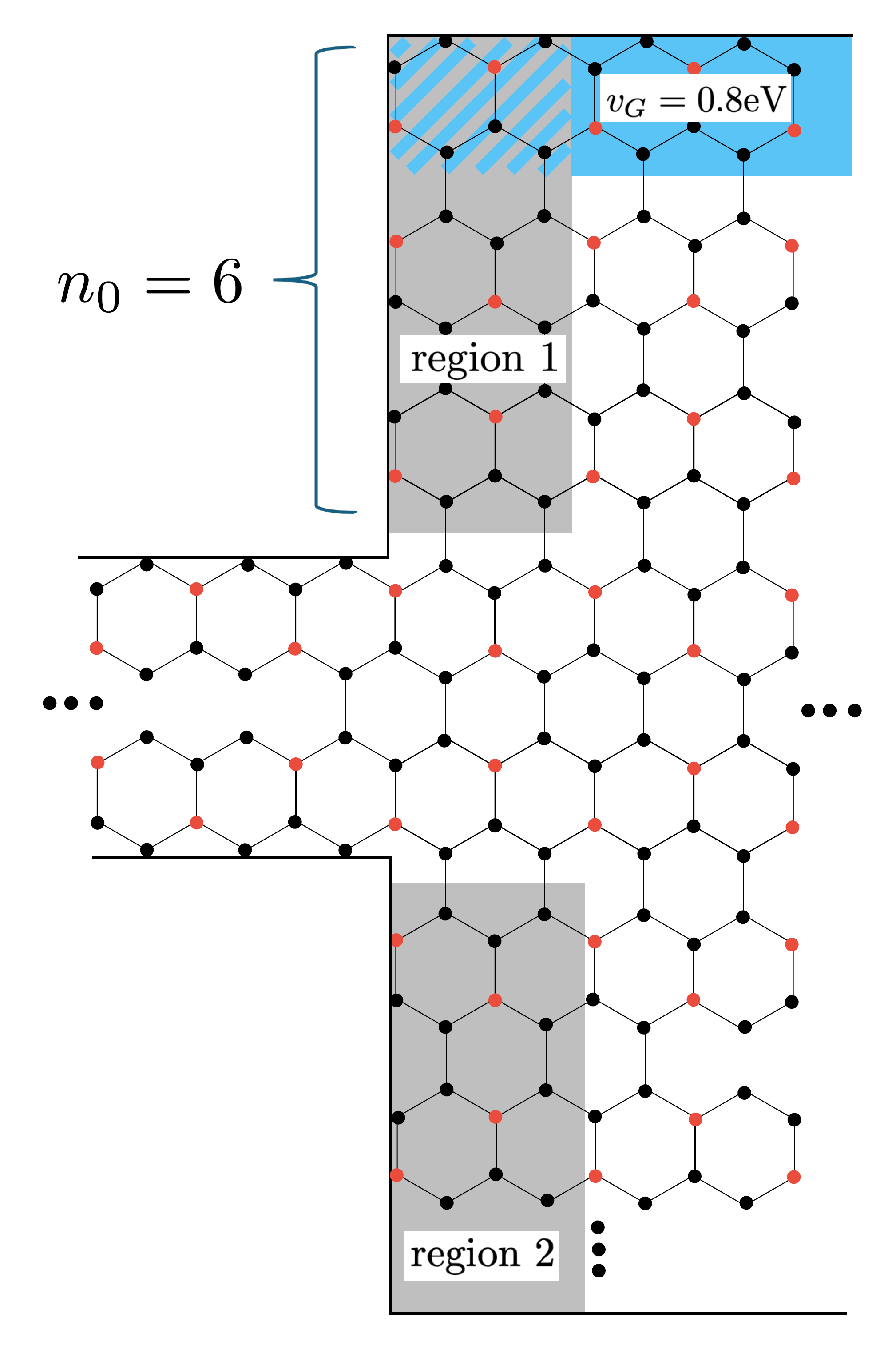}
    \caption{Mismatched nanojunction formed by a narrow $N_T=4$ semi-infinite ZNR joined with a wide $N_T=50$ ZNR. The mismatch $n_0$ of the upper edge of the junction counts the number of chains one needs to add on the upper edge of the left lead to align with the upper edge of the right lead. For the right lead, a gate potential of $v_G=0.8$ eV is added to all atomic onsite potentials of the upper two chains.}
    \label{fig:ribbonmissmatch}
\end{figure}

We investigated mismatched nanojunctions formed by a semi-infinite ZNR with $N_T=50$ with C-C edge termination attached to a narrow semi-infinite ZNR with $N_T=4$. 
The two semi-infinite ribbons are not necessarily aligned at one of the edges, the mismatch is denoted as $n_0$ and is determined by the number of chains between the two upper edges of the left and right semi-infinite ZNR as illustrated in Fig.~\ref{fig:ribbonmissmatch}. The atomic configuration is chosen such that at the interface between left and right ZNR there are no defects, i.e., for the chains contacted through the interface, the atomic configuration is as in an ideal ribbon. For the right ZNR we also apply the gate potential $v_G=0.8$ eV for all atoms of the first two chains (see Fig.~\ref{fig:ribbonmissmatch}). 

It is worth noting that the ribbon widths selected here serve primarily to illustrate the proposed design. The desired effect, i.e., the formation of Fano resonances, is a general feature which can be observed in a variety of different structures, as long as one ribbon is significantly narrower than the other one and also exhibits states in the desired energy range. Similarly, the gate potential may take a range of values or may be applied to not only the first two chains of atoms. While specific numerical results may vary depending on the chosen parameters, the qualitative behavior remains highly robust. The fundamental mechanism for Fano resonance formation is the energetic shift of edge-state bands into the region of energies of the interface states.

In Fig.~\ref{fig:LDOSregion1and2} we show the (local) densities of states $D_{C1 (C2)}(E) $ of regions 1 and 2 marked in Fig.~\ref{fig:ribbonmissmatch} for the mismatched nanojunction formed by two ribbons of width $N_T=4$ and $N_T=50$ with a mismatch of $n_0=12$. We see that for region 1, $D_{C1}$ is continuous but exhibits resonance features with the typical shape of Fano resonances (see discussion below). In contrast, in region 2, $D_{C2}(E) $  shows a number of sharp peaks indicating the presence of localized states. The clear difference between $D_{C1}$ and $D_{C2}$ can be understood from the modified geometry of the mismatched junction as compared to the semi-infinite ZNR. In the latter structure, the corresponding surface states exist along the entire interface. As a result, once the energy band corresponding to the upper edge is shifted into the same energy range, its hybridization with the interface states is visible over the full interface.
However, for the mismatched junction this continuous interface is effectively divided into two spatially separated segments corresponding to regions 1 and 2. Each segment can now create its own localized interface states. Since only the upper segment (region 1) remains strongly coupled to the upper edge band state of the right ribbon, the hybridization with the continuum is mainly observed in region 1. On the other hand, the DOS $D_{C2}$ for region 2 exhibits sharp peaks with only a small finite linewidth. This indicates that the corresponding states are not completely isolated and remain weakly coupled to the rest of the system. 
Since the states in the two interface segments are not strictly independent, the discrete peaks associated with region 2 can still have some spectral weight in $D_{C1}$, as can be clearly seen in Fig.~\ref{fig:LDOSregion1and2} a).

\subsection{Fano resonances through interference of edge band and localized interface states}

\begin{figure}[tb]
    \includegraphics[width=1.0\linewidth]{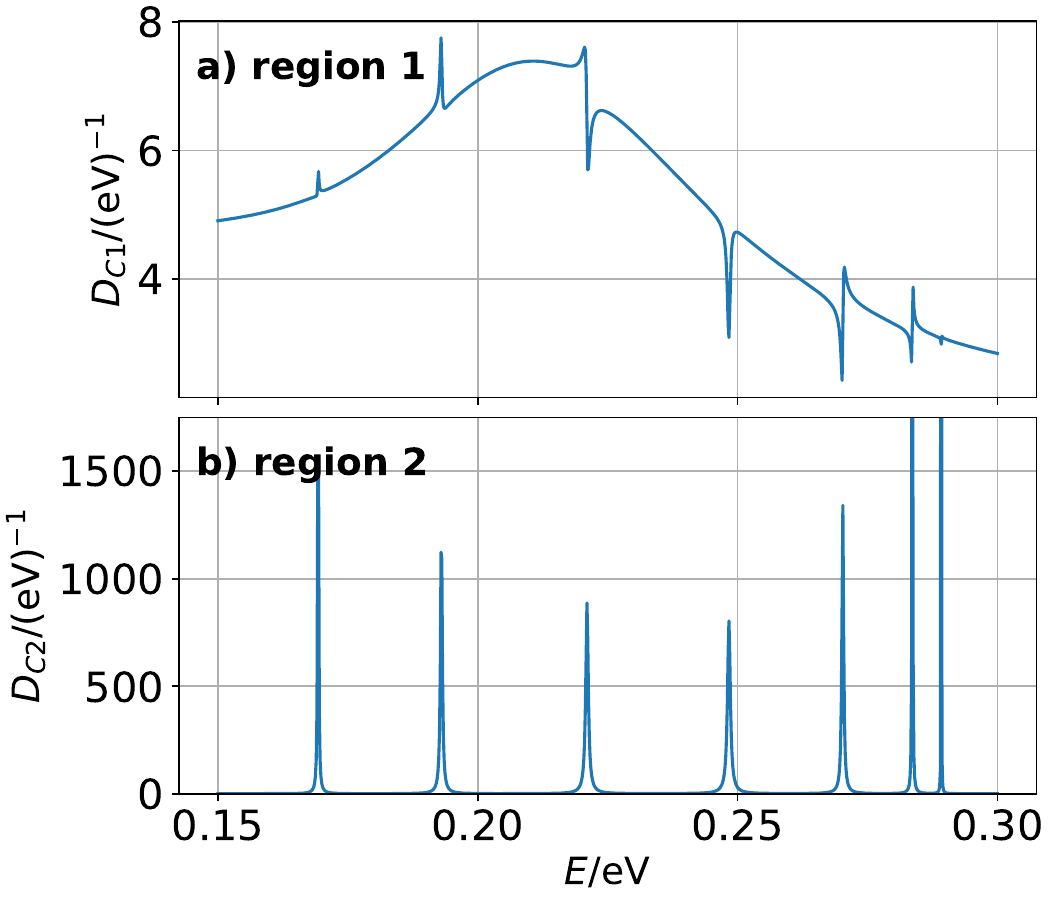}
    \caption{Density of states for region 1, $D_{C1}$, (upper panel) and region 2, $D_{C2}$, (lower panel) for a mismatched nanojunction. For the definition of regions 1 and 2, see Fig.~\ref{fig:ribbonmissmatch}. The junction is formed by two semi-infinite ZNRs with widths $N_T=4$ and $N_T=50$, respectively, which are  attached to each other with a mismatch $n_0=12$. For the right ZNR, a gate potential of $v_G=0.8$ eV has been applied to all onsite energies of chains $N=1$ and $N=2$.}
    \label{fig:LDOSregion1and2}
\end{figure}

The line shapes of the peaks in $D_{C1}$ in Fig.~\ref{fig:LDOSregion1and2} suggest the formation of Fano resonances, an  effect which can be understood as follows: attaching the left ribbon to the right one leads to a modification of the interface geometry as well as the localized interface states. The interface states in region 1 strongly couple to the edge state localized at the upper edge of the right ribbon, and, together with those of the left ribbon, form an open transport channel. If the weakly broadened (quasi-)discrete localized states in region 2 (see Fig.~\ref{fig:LDOSregion1and2} b) lie in the same energy range as this open channel, the two types of states can interfere.
This situation is characteristic for the appearance of Fano resonances, which arise from the interference between a (quasi-)discrete state and a continuum of propagating states~\cite{Fano:61}. 
As a consequence, Fano resonances are expected to be present in our mismatched junction.

\begin{figure}[tb]  \includegraphics[width=1\linewidth]{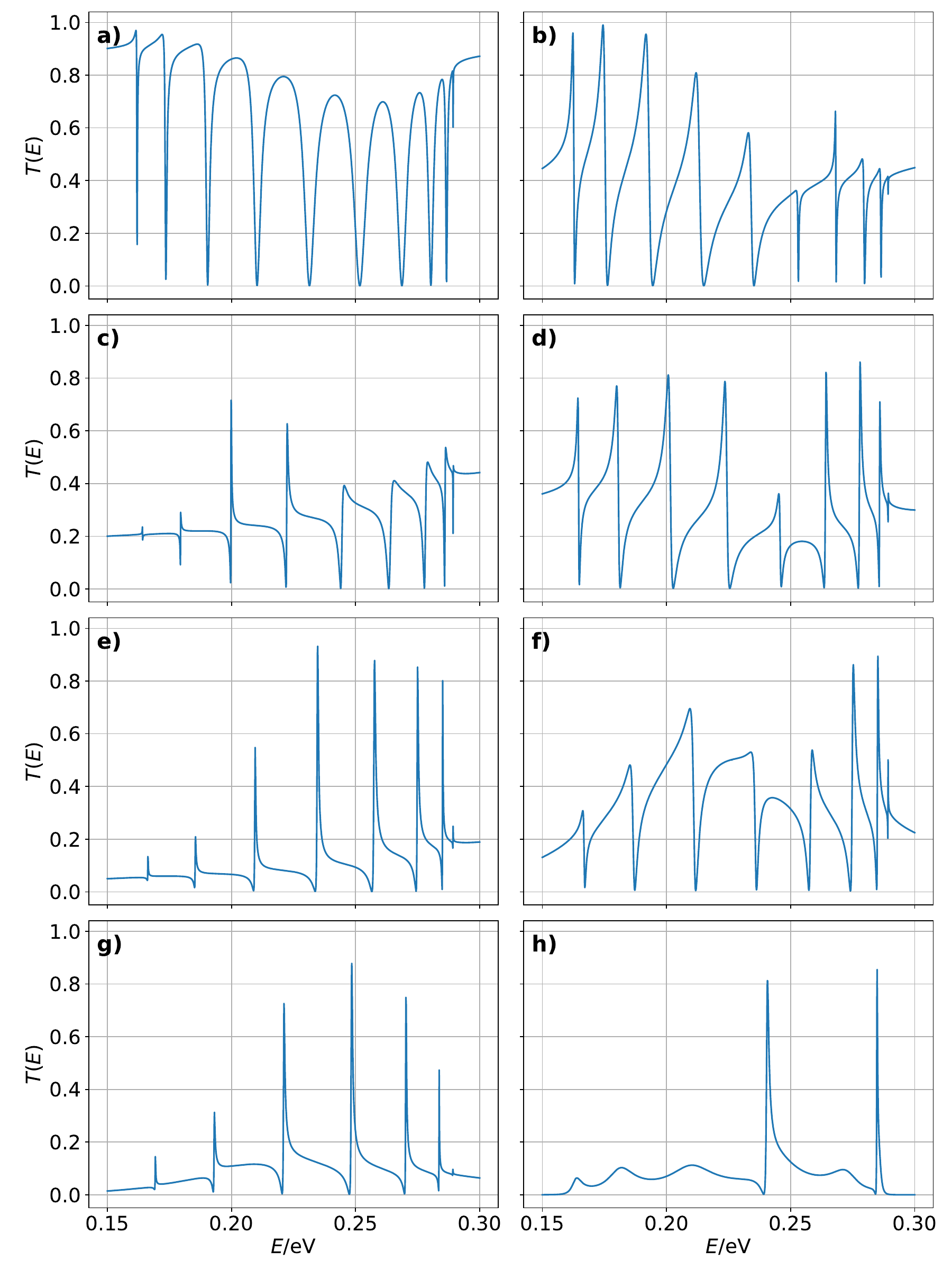}
    \caption{Transmission functions of mismatched nanojunctions for different values $n_0$ of the mismatch. Panels  a) $n_0=0$ b) $n_0=2$, c) $n_0=4$, d) $n_0=6$, e) $n_0=8$, f) $n_0=10$, g) $n_0=12$, h) $n_0=32$. }
    \label{fig:fanotrans}
\end{figure}

Fano resonances are present not only in $D_{C1}$, but can also be observed in the transmission function. In Fig.~\ref{fig:fanotrans} we show the transmission functions of several mismatched nanojunctions for different values of the mismatch $n_0$. The other parameters remain the same as in the previous setup, i.e, the widths of the left and right ribbons are $N_T=4$ and $N_T=50$, respectively, and the gate potential of $v_G=0.8$ eV is applied to all atoms on chains $1$ and $2$. For most of the values of the mismatch $n_0$, we clearly see Fano resonances in the transmission. For large values of $n_0$, however, the number of Fano resonances is significantly reduced. This is simply due to the fact that if one increases $n_0$, one reduces the size of region 2 and decreases the number of discrete peaks (states), which can contribute to the formation of a Fano resonance.

Another interesting observation is the difference in the shape of the Fano resonances for $n_0/2$ even (panels a), c), e), g), and h) of Fig.~\ref{fig:fanotrans}) and $n_0/2$ odd (panels b), d), and f) of Fig.~\ref{fig:fanotrans}). For $n_0/2$ even, the Fano resonances always have the shape of a sharp rise followed
by a less sharp shoulder as the energy increases. On the other hand, for $n_0/2$ odd, the Fano resonances at lower energies start with a shoulder followed by peak with very sharp decrease. However, at some point in the energy range the shape of the resonances is reversed, i.e., a sharp peak is followed by a shoulder as energy increases. 
We attribute this different behavior between even and odd $n_0/2$ to the different local atomic structure right at the interface between the $N_T=4$ and the
$N_T=50$ ZNR.

The transmission function near a Fano resonance can be written in terms of the Fano representation
\begin{equation}
    T(E)=a F(\epsilon)+T_{bg} 
    %\text{   ; } \epsilon=\frac{E-E_0}{\Gamma/2}
    \label{fano_lineshape}
\end{equation}
with 
\begin{equation}
    \epsilon= \frac{E-E_0}{\Gamma/2}
\end{equation}
where $E_0$ and $\Gamma$ are the position and the half-width of the resonance, while 
\begin{equation}
    F(\epsilon) = \frac{(\epsilon+q)^2}{(1+q^2)(1+\epsilon^2)}
\end{equation}
is the normalized Fano resonance profile (whose values are restricted to the interval $[0,1]$)\cite{MiroshnichenkoFlachKivshar:10}. Here, $a$ is a scaling factor and $q$ is the Fano asymmetry parameter which determines the asymmetry of the Fano lineshape. Finally, $T_{bg}$ is the ``background" contribution to the transmission, here taken to be an energy-independent parameter.

\begin{figure}[bt]
\includegraphics[width=1\linewidth]{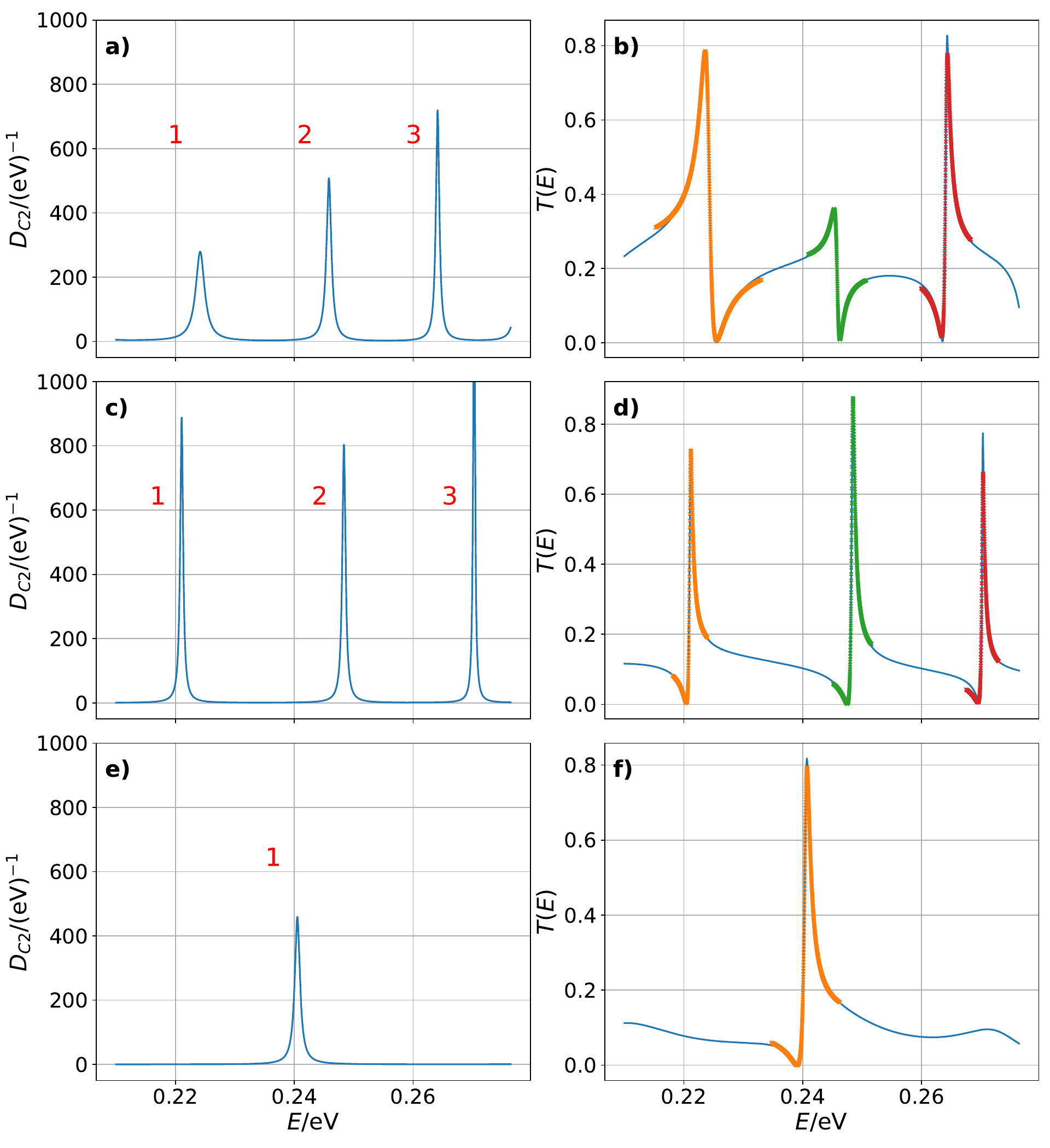}
    \caption{$D_{C2}$ (left panels) and transmission functions (right panels) of mismatched nanojunctions between a (semi-infinite) $N_T=4$ and a $N_T=50$ ZNR for different offsets $n_0$, a) and b) $n_0=6$, c) and d) $n_0=12$, e) and f) $n_0=32$. In the right panels, the fits to the Fano resonances according to Eq.~(\ref{fano_lineshape}) are included. The corresponding fit parameters are listed in Table \ref{tab:parameters}.}
    \label{fig:FANOFIT}
\end{figure}

In Fig.~\ref{fig:FANOFIT} we show the DOS of region 2, $D_{C2}$, (left panels) and the  transmission functions (right panels) for three different offsets $n_0$ together with the fits of the Fano resonances to Eq.~(\ref{fano_lineshape}). First we determine both the position $E_0$ and the half-width $\Gamma$ of a resonance from $D_{C2}$. With these parameters fixed, we then fit the other parameters ($a$, $q$, and $T_{bg}$) of the Fano lineshapes. 
The values of the parameters for the different resonances are given in Table \ref{tab:parameters}. We see that all the Fano resonances are reproduced extremely well with the fits according to Eq.~(\ref{fano_lineshape}).

\begin{table}[th]
  \centering
    \begin{tabular}{|c|c|c|c|c|c|c|}\hline 
    $n_0$ & Peak & $E$/eV & $\Gamma$/eV & $q$ & $a$ & $T_{bg}$ \\ \hline
    & 1 & 0.224& 0.00178& -1.541& 0.777& 0.007\\  
  6  & 2 & 0.246& 0.00098&  -0.897& 0.350&0.009\\  
    & 3 & 0.264& 0.00084& 1.724& 0.757&0.018\\ \hline
    & 1 & 0.221& 0.00058& 2.157& 0.721& 0.004\\
  12  & 2 & 0.248& 0.00064& 2.681& 0.873& 0.002\\    
    & 3 & 0.270& 0.00054& 2.843& 0.654& 0.005\\ \hline 
  32  & 1 & 0.240& 0.00148& 2.519& 0.793&0.000\\ \hline
    \end{tabular}
    \caption{Values of the fit parameters for the different Fano resonances of Fig.~\ref{fig:FANOFIT}. }
    \label{tab:parameters}
\end{table}

\section{Conclusions}
\label{sec:conclusions}

In this work we have shown how one can engineer Fano resonances in the transmission functions of mismatched nanojunctions formed by two polyaniline (C$_3$N) zigzag nanoribbons of different widths. We have investigated the electronic properties of these mismatched nanojunctions by using a tight-binding model in combination with the non-equilibrium Green's function technique. The engineering of the desired properties has been achieved using several steps: (i) we took advantage of the presence of two bands of edge states in the ideal nanoribbon. By applying a constant gate potential to the atoms in the vicinity of {\em one} edge of the ribbon, the range of energies of only one of the edge bands may be manipulated at will; (ii) for a semi-infinite nanoribbon, localized states at the vertical armchair interface are formed. These localized interface states can hybridize with the edge states if the edge-state band is shifted with a gate potential to energies accessible to the interface states; (iii) attaching to a semi-infinite nanoribbon another one of different width (and thus forming a mismatched nanojunction) may lead  to the formation of an open transport channel along the ribbon edge. At the same time, the mismatched nanojunction may again allow for the formation of interface states, which are now localized at the side of the mismatched junction opposite to the open edge channel. Therefore, we have created a setup where electrons from the continuum of edge states can interfere with localized interface states, a prototypical scenario for the formation of Fano resonances. These Fano resonances can be observed both in the density of states (in spatial regions where the open transport channel is located) as well as in the transmission profiles. The shape of these resonances in the transmission can be described perfectly with the canonical form of the Fano profile.  

While here we have studied the formation of Fano resonances for
  C$_3$N mismatched ZNR nanojunctions (in a 
TB model), the effect as such should be rather general. It requires the existence of localized states (e.g., at the interface 
of mismatched nanojunctions) which are energetically close to continuum states such that they can hybridize. If this 
happens close to the Fermi energy, the effect may possibly even be exploited for new electronic devices based on carefully 
designed nanojunctions formed by C$_3$N or other materials.

The experimental realization of devices exploiting Fano resonances requires
precise control over sample quality and gate-voltage gating, a common
challenge in nanoelectronics.  Recent top-down and bottom-up lithographic
methods have achieved sub-nanometer precision, e.g., graphene nanoribbons with
widths smaller than one nanometer have been successfully synthesized in
ultrahigh vacuum environments \cite{Kimouche2015}. Given the structural
similarities, such high-precision fabrication frameworks can be extended
to C$_3$N nanoribbons. Therefore, the integration of the proposed mismatched
nanoribbons into functional nanoelectronic architectures, such as field-effect
transistors, is entirely within reach of current experimental capabilities.

\section*{Acknowledgements}

A.C.P. acknowledges fruitful discussions with Carsten Timm. 
R.D'A. kindly acknowledges the hospitality of the Physics Department of the University of Milan ``Statale" where some of this work has been completed. 
S.K. acknowledges financial support from the Basque Government through the Elkartek program (project CICe2025, grant number KK2025-00054).

\end{document}